\documentclass{article}
\usepackage[final,nonatbib]{nips_2017}
\usepackage{graphicx}				% Use pdf, png, jpg, or eps§ with pdflatex; use eps in DVI mode
\usepackage{float}
\usepackage{listings}
\usepackage{pygtex}
\usepackage{amssymb}
\usepackage{hyperref}
\usepackage[utf8]{inputenc} % allow utf-8 input
\usepackage[T1]{fontenc}    % use 8-bit T1 fonts
\usepackage{url}            % simple URL typesetting
\usepackage{booktabs}       % professional-quality tables
\usepackage{amsfonts}       % blackboard math symbols
\usepackage{nicefrac}       % compact symbols for 1/2, etc.
\usepackage{microtype}      % microtypography
\usepackage{xpatch,letltxmacro}
\usepackage{multicol}
\usepackage[compatibility=false]{caption}
\usepackage{subcaption}

\newfloat{listing}{tbhp}{lst}
\floatname{listing}{Listing}

\newcommand{\TODO}[1]{}

\title{Tangent: Automatic Differentiation Using Source Code Transformation in Python}
\author{
    Bart van Merriënboer\\
    Google Inc.\\
    \texttt{bartvm@google.com}\\
    \And
    Alexander B Wiltschko\\
    Google Inc.\\
    \texttt{alexbw@google.com}
    \And
    Dan Moldovan\\
    Google Inc.\\
    \texttt{mdan@google.com}\\
}

\begin{document}
\maketitle

\begin{abstract}
\emph{Automatic differentiation} (AD) is an essential primitive for machine learning programming systems.
Tangent is a new library that performs AD using \emph{source code
transformation} (SCT) in Python. It takes numeric functions written in a
syntactic subset of Python and NumPy as input, and generates new
Python functions which calculate a derivative. This approach to automatic
differentiation is different from existing packages popular in machine
learning, such as TensorFlow\cite{abadi2016tensorflow} and
Autograd\footnote{\url{https://github.com/HIPS/autograd}}. Advantages are that
Tangent generates gradient code in Python which is readable by the user, easy to understand and debug, and has no runtime overhead. Tangent also introduces abstractions for easily injecting logic into the generated gradient code, further improving usability.
\end{abstract}

%% Be more explicit and clear about _advantages_:
%% 1. AD applies to pure, real Python, not a weird custom dataflow graph - vs. TF/Theano
%% 2. Gradient computation is performed by a simple, standaldone pure Python function, like a programmer might write by hand, but automatically generated - straightforward debugging and introspection vs. Autograd/PyTorch "tape"
%% 3. Gradient transformation can be easily overloaded/customized, both for expressiveness (custom approximations) and usability (logging, debug hook points, ...) - vs. Autograd/PyTorch, Theano - TF arguably has this if you just customize your backprop implementation for a given node type, right?

\section{Introduction}

Modern machine learning and deep learning relies heavily on gradient-based optimization algorithms. These methods require the efficient calculation of derivatives of potentially complex, high-dimensional models. AD is a set of techniques to evaluate these derivatives. It is based on the insight that the chain rule can be applied to the elementary arithmetic operations (primitives) performed by a program, and nearly every machine learning library implements it. Note that AD is different from symbolic differentiation (which applies to mathematical expressions instead of programs) and numerical differentiation (where the gradient is approximated using finite differences).

Two approaches to automatic differentiation are common: \emph{tracing} and \emph{source code transformation} (SCT). In the tracing approach, primitives are overloaded so that each operation is logged onto
a tape (a linear trace) at runtime. The chain rule can then be applied by walking this tape backward. Source code transformation, on the other hand, explicitly rewrites the code prior to execution to produce a separate gradient version. Both approaches have different implementation, performance, and usability trade-offs \cite{bischof2000computing}.

Automatic differentiation packages using both approaches have long existed for, e.g., C, C++, Fortran, (see \cite{baydin2015automatic} for an overview) and have been used in fields such as computational fluid dynamics, atmospheric sciences, and astronomy. The machine learning community's different needs, which include a heavy focus on linear algebra, performance through heterogenous (GPGPU) computing, and a pervasive use of Python, led to the development of separate tools.

Theano \cite{al2016theano} and TensorFlow \cite{abadi2016tensorflow} are two popular machine learning
frameworks with support for AD. Although Python-based, they do not perform AD
on the Python code. Instead, Python is used as a metaprogramming language to
define a dataflow graph (computation graph) on which SCT is performed. Since
these dataflow graphs do not have function calls or lexical scoping, the AD
logic is simplified. However, the introduction of a separate programming
paradigm which requires its own runtime can be confusing to the user.

AD has been implemented for Python and NumPy using tracing in the Autograd and ad\footnote{\url{http://pythonhosted.org/ad/}} packages\footnote{A more complete list of AD tools in several languages: \url{http://www.autodiff.org/?module=Tools}}.

\section{Benefits of Source Code Transformation for Automatic Differentiation}

Because Tangent is (to our knowledge) the first SCT-Based AD system for Python, it occupies a unique point in the space of tradeoffs among usability, flexibilty, debuggability, and computational performance. It allows different tradeoffs in usability, and ease of debugging than prior systems.

\subsection{Usability}

The metaprogramming approach used by Theano and Tensorflow often results in more verbose and less idiomatic code (see Listing \ref{lst:listingxx}, left).

\begin{figure}[!h]
 \begin{minipage}[t][3.5cm][t]{0.55\textwidth}
  \centering
  \input{1-py}
 \end{minipage}
 \begin{minipage}[t][3cm][t]{0.45\textwidth}
  \centering
  \input{2-py}
 \end{minipage}
 \captionof{listing}{Left: Gradient of $x\cdot x$ in TensorFlow. Right: Gradient of $x\cdot x$ in Tangent and Autograd, which both use a similar API: \texttt{grad(f)}}
  \label{lst:listingxx}
\end{figure}
Autograd and Tangent allow users to write models in more idiomatic code (see Listing \ref{lst:listingxx}, right).
This is particularly useful for code that contains control flow constructs (see Listing \ref{lst:loop})

\begin{figure}[!h]
 \begin{minipage}[t][6cm][t]{0.55\textwidth}
  \centering
  \input{3-py}
 \end{minipage}
 \begin{minipage}[t][3cm][t]{0.45\textwidth}
  \centering
  \input{4-py}
 \end{minipage}
 \captionof{listing}{Left: Implementation of a differentiable nested loop and conditional in TensorFlow. Right: Implementation of the same program in Tangent. Much less code is required, due to the use of native Python syntax.}
  \label{lst:loop}
\end{figure}

However, the tracing approach can be problematic for debugging and usability. When the function \texttt{df} is called, the function \texttt{f} is executed with non-standard semantics (logging to the tape), after which the tape is walked in reverse using a loop that is internal to Autograd. Errors that occur anywhere during execution will potentially have tracebacks that are hard to understand for the user, because they are buried inside the Autograd implementation.

Because Tangent uses source code transformation, the function \texttt{df} that Tangent generates is a new Python function, with standard semantics, whose source code can be directly inspected (see Listing \ref{lst:tangentxx}). This simplifies both user understanding and debugging.

\begin{listing}
\centering
  \begin{Verbatim}[commandchars=\\\{\},frame=none,framesep=1.5ex,framerule=0.8pt]
\PY{k}{def} \PY{n+nf}{dfdx}\PY{p}{(}\PY{n}{x}\PY{p}{,} \PY{n}{by}\PY{o}{=}\PY{l+m+mf}{1.0}\PY{p}{)}\PY{p}{:}
    \PY{c+c1}{\PYZsh{} Grad of: y = x * x}
    \PY{n}{\PYZus{}bx} \PY{o}{=} \PY{n}{tangent}\PY{o}{.}\PY{n}{unbroadcast}\PY{p}{(}\PY{n}{by} \PY{o}{*} \PY{n}{x}\PY{p}{,} \PY{n}{x}\PY{p}{)}
    \PY{n}{\PYZus{}bx2} \PY{o}{=} \PY{n}{tangent}\PY{o}{.}\PY{n}{unbroadcast}\PY{p}{(}\PY{n}{by} \PY{o}{*} \PY{n}{x}\PY{p}{,} \PY{n}{x}\PY{p}{)}
    \PY{n}{bx} \PY{o}{=} \PY{n}{\PYZus{}bx}
    \PY{n}{bx} \PY{o}{=} \PY{n}{tangent}\PY{o}{.}\PY{n}{add\PYZus{}grad}\PY{p}{(}\PY{n}{bx}\PY{p}{,} \PY{n}{\PYZus{}bx2}\PY{p}{)}
    \PY{k}{return} \PY{n}{bx}
\end{Verbatim}

\caption{Source code of gradient of $x\cdot x$ in Tangent. The \texttt{unbroadcast} is responsible for reversing the broadcasting done by NumPy when performing element-wise operations on differently-sized multidimensional arrays.}
\label{lst:tangentxx}
\end{listing}

%\subsection{Python ecosystem}
%
%In general, the usability improvements that Tangent provides stem from the fact that it generates explicit Python code for the gradient. In general, this has the advantage that gradient code can be processed seamlessly by the rest of the Python ecosystem. For example, it can be compiled using tools such as Cython and Numba, and profiled using standard tools such as \texttt{line\_profiler} and \texttt{memory\_profiler}.

\subsection{Injecting Custom Logic Into Gradients}
There are several cases in which it can be useful for the user to inject custom code into the gradient computation.

Many algorithms use approximations or modifications of the gradient. For example, for performance reasons, recurrent neural networks (RNNs) are often trained using truncated backpropagation through time~\cite{williams1990efficient} (TBPTT). This algorithm  performs fewer loop iterations in the gradient than in the original function. In other cases, custom gradients are used to train models with discontinuous functions (e.g. using straight-through estimators~\cite{bengio2013estimating}).

Second, debugging errors in the gradient computation (e.g., overflow/underflow) can greatly benefit from the ability to insert arbitrary code into the generated gradient, which allows the user to add logging statements or insert breakpoints.

Traditional AD frameworks have little support for this kind of code injection. Theano and TensorFlow allow the user to manipulate the dataflow graph of the gradient directly to accomplish some of these changes, but this can be cumbersome. Tangent overloads Python's context manager syntax to introduce a novel way of allowing the user to inject arbitrary code into the gradient computation. We have found this syntax to be a very succinct way of implementing these cases.

\begin{listing}
\centering
  \begin{Verbatim}[commandchars=\\\{\},frame=none,framesep=1.5ex,framerule=0.8pt]
\PY{k}{def} \PY{n+nf}{f}\PY{p}{(}\PY{n}{x}\PY{p}{)}\PY{p}{:}
    \PY{k}{with} \PY{n}{grad\PYZus{}of}\PY{p}{(}\PY{n}{x}\PY{p}{)} \PY{k}{as} \PY{n}{dx}\PY{p}{:}
        \PY{k}{if} \PY{n}{dx} \PY{o}{\PYZgt{}} \PY{l+m+mi}{10}\PY{p}{:}
            \PY{k}{print}\PY{p}{(}\PY{l+s+s1}{\PYZsq{}}\PY{l+s+s1}{Warning, large gradient of x}\PY{l+s+s1}{\PYZsq{}}\PY{p}{,} \PY{n}{dx}\PY{p}{)}
        \PY{n}{dx} \PY{o}{/}\PY{o}{=} \PY{l+m+mi}{2}
    \PY{k}{return} \PY{n}{x} \PY{o}{*} \PY{n}{x}
\end{Verbatim}

\caption{Gradient manipulation; in this case \texttt{df(2)} will return 2, because the gradient is halved. Logging statements for large gradient values (or \texttt{NaN} gradient values) are also easily inserted.}
\label{lst:gradof}
\end{listing}

\section{Implementation}

% Python's standard library includes many of the necessary machinery to implement AD. The \texttt{inspect} library gives us access to the source code of any user-defined function. The \texttt{ast} library allows us to parse this source code into an abstract syntax tree (AST), on which we can perform source code transformation. We use the Astor\footnote{\url{https://github.com/berkerpeksag/astor}} package to transform the AST into plain text again, which is then executed by the Python interpreter as usual.

Tangent uses Python's built-in machinery to introspect and transform the \emph{abstract syntax tree} (AST) of parsed source code at runtime. For each piece of supported Python syntax, we have implemented a rule indicating how to rewrite an AST node into its backward pass equivalent, or "adjoint". We have defined adjoints for function calls to NumPy methods, as well as larger pieces of syntax, such as if-statements and for-loops. The adjoints are stored in function definitions that serve as "templates", or code macros \cite{macro}. Another alternative, which we found too cumbersome, would be to use a templating engine like Mustache \footnote{\url{https://mustache.github.io/}} and store adjoints as plain strings. Our templates also use a special syntax \texttt{d[x]} to refer to the derivative of a variable \texttt{x} (see Listing \ref{lst:quoting}).

\begin{listing}
\centering
  \begin{Verbatim}[commandchars=\\\{\},frame=none,framesep=1.5ex,framerule=0.8pt]
\PY{c+c1}{\PYZsh{} Code quote specifying the gradient of np.multiply.}
\PY{c+c1}{\PYZsh{} This function serves only as a container for code that will be}
\PY{c+c1}{\PYZsh{} expanded and in\PYZhy{}lined in generated code.}
\PY{k}{def} \PY{n+nf}{adjoint\PYZus{}multiply}\PY{p}{(}\PY{n}{result}\PY{p}{,} \PY{n}{arg1}\PY{p}{,} \PY{n}{arg2}\PY{p}{)}\PY{p}{:}
  \PY{n}{d}\PY{p}{[}\PY{n}{arg1}\PY{p}{]} \PY{o}{=} \PY{n}{arg2} \PY{o}{*} \PY{n}{d}\PY{p}{[}\PY{n}{result}\PY{p}{]}
  \PY{n}{d}\PY{p}{[}\PY{n}{arg2}\PY{p}{]} \PY{o}{=} \PY{n}{arg1} \PY{o}{*} \PY{n}{d}\PY{p}{[}\PY{n}{result}\PY{p}{]}


\PY{c+c1}{\PYZsh{} To generate the adjoint for this line...}
\PY{n}{var3} \PY{o}{=} \PY{n}{np}\PY{o}{.}\PY{n}{multiply}\PY{p}{(}\PY{n}{var1}\PY{p}{,}\PY{n}{var2}\PY{p}{)}

\PY{c+c1}{\PYZsh{}  ... we use macro expansion}
\PY{n}{new\PYZus{}ast} \PY{o}{=} \PY{n}{tangent}\PY{o}{.}\PY{n}{template}\PY{o}{.}\PY{n}{replace}\PY{p}{(}\PY{n}{adjoint\PYZus{}multiply}\PY{p}{,}
                                   \PY{n}{result}\PY{o}{=}\PY{l+s+s1}{\PYZsq{}}\PY{l+s+s1}{var3}\PY{l+s+s1}{\PYZsq{}}\PY{p}{,}
                                   \PY{n}{arg1}\PY{o}{=}\PY{l+s+s1}{\PYZsq{}}\PY{l+s+s1}{var1}\PY{l+s+s1}{\PYZsq{}}\PY{p}{,}
                                   \PY{n}{arg2}\PY{o}{=}\PY{l+s+s1}{\PYZsq{}}\PY{l+s+s1}{var2}\PY{l+s+s1}{\PYZsq{}}\PY{p}{)}

\PY{c+c1}{\PYZsh{} If the code specified in the AST \PYZsq{}new\PYZus{}ast\PYZsq{} were converted into a string:}
\PY{n}{b\PYZus{}var1} \PY{o}{=} \PY{n}{var2} \PY{o}{*} \PY{n}{b\PYZus{}var3}
\PY{n}{b\PYZus{}var2} \PY{o}{=} \PY{n}{var1} \PY{o}{*} \PY{n}{b\PYZus{}var3}
\end{Verbatim}

\caption{Underlying implementation of gradient code construction. The user will
not routinely write code in this style, unless implementing custom gradients.
The gradient of \texttt{np.multiply} is specified as a code quote. This
function will never be run as-is --- it only contains a template that will be
used to generate contextually-correct gradient code. Tangent internally discovers
the names of the arguments and results in the code snippet and looks up the appropriate
template (that process is omitted here). Then, Tangent combines the template and
the variable names to create a new AST.}

\label{lst:quoting}
\end{listing}

Tangent is restricted to a subset of Python where functions have no side effects. Mutating arrays is allowed, but only through index assignment syntax (\texttt{a[i] = b}). This requirement prevents us from having to copy large multi-dimensional arrays each time they are used. Tangent also does not support closures, because closures with free variable references lead to a problem sometimes referred to as `perturbation confusion', which is non-trivial to address~\cite{Pearlmutter}.
These restrictions only apply to statements that involve active variables i.e.,
variables which affect the output of the function whose derivative we are
computing.

Two other limitations on the supported subset of Python are worth mentioning. First, if the name of a function cannot be tracked to its source code ahead-of-time, we will generate an error. This situation may arise if functions are being passed as variables, or if they are being renamed in the user's code. A second case is the use of classes, and class member functions. This is an important use case, as many large neural network models are coded using an object-oriented style. Tangent does not currently support taking derivatives through classes, although we are actively working on this feature.

We optimize generated code for both readability and performance. We do this by constructing a control flow graph (CFG) from the AST in order to determine which variables are active (a form of forward dataflow analysis). To improve readability of the final code and performance, we use Tangent's ability to perform dataflow analysis on Python code to perform several simplifications on the transformed AST (similar to an optimizing compiler). We perform algebraic simplifications e.g., instead of explicitly initializing a gradient to zero and accumulating into it (\texttt{dx = 0; dx += 2}) we simply assign (\texttt{dx = 2}) where possible. We also perform dead code elimination (cf. Listing \ref{lst:tangentxx} where the original statement, \texttt{y = x * x}, was removed).

\section{Performance}

Because Tangent performs AD ahead-of-time, it has no runtime overhead. Its performance then depends largely on the CPython interpreter and implementations of the underlying numeric kernels, such as matrix multiplication and convolution. Here, we compare the performance of Tangent, TensorFlow and Autograd, a tracing automatic differentiation system for NumPy. As a simple benchmark, we use a multi-layer perceptron, implemented in NumPy (see Listing \ref{lst:mlp}).

\begin{figure}
\begin{subfigure}{.5\textwidth}
  \centering
  \includegraphics[width=.8\linewidth]{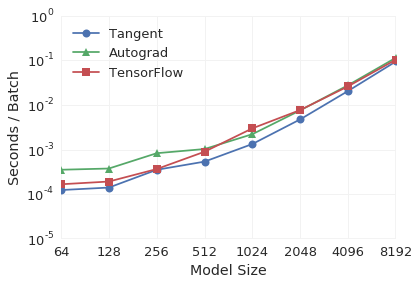}
  \caption{Benchmark results for the MLP from Listing \ref{lst:mlp}.}
  \label{fig:benchmark-mlp}
\end{subfigure}%
\begin{subfigure}{.5\textwidth}
  \centering
  \includegraphics[width=.8\linewidth]{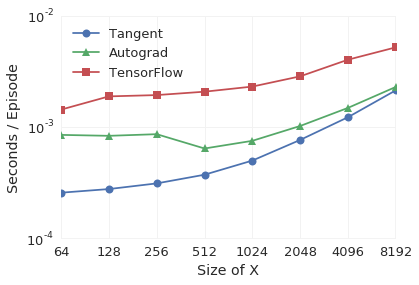}
  \caption{Benchmark for a simple loop from Listing \ref{lst:loop}.}
  \label{fig:benchmark-loop}
\end{subfigure}
\caption{Average of 50 runs, batch size of 16, for varying number of parameters. Run on a Xeon E5-1650 v3 @ 3.5 GHz, 64GB of RAM, with Ubuntu 14.04 on Python 2.7, with MKL.}
\label{fig:fig}
\end{figure}

Tangent's performance is superior to Autograd on a simple multi-layer perceptron benchmark (Figure \ref{fig:benchmark-mlp}), particularly for smaller model sizes, where overhead dominates. Autograd\footnote{Code checked out from: \url{https://github.com/HIPS/autograd/tree/7fa48ab4c}} adds interpretive overhead at every gradient call, because it first traces user code, and then interprets the trace to calculate the derivative. Tangent largely matches the performance of TensorFlow, slightly exceeding it for some model sizes. Tangent is faster for all model sizes for the simple loop benchmark (Figure \ref{fig:benchmark-loop}).

\begin{listing}
\centering
  \begin{Verbatim}[commandchars=\\\{\},frame=none,framesep=1.5ex,framerule=0.8pt]
\PY{k}{def} \PY{n+nf}{logsumexp}\PY{p}{(}\PY{n}{x}\PY{p}{,} \PY{n}{axis}\PY{o}{=}\PY{n+nb+bp}{None}\PY{p}{,} \PY{n}{keep\PYZus{}dims}\PY{o}{=}\PY{n+nb+bp}{False}\PY{p}{)}\PY{p}{:}
  \PY{k}{return} \PY{n}{np}\PY{o}{.}\PY{n}{log}\PY{p}{(}\PY{n}{np}\PY{o}{.}\PY{n}{sum}\PY{p}{(}\PY{n}{np}\PY{o}{.}\PY{n}{exp}\PY{p}{(}\PY{n}{x}\PY{p}{)}\PY{p}{,} \PY{n}{axis}\PY{o}{=}\PY{n}{axis}\PY{p}{,} \PY{n}{keepdims}\PY{o}{=}\PY{n}{keep\PYZus{}dims}\PY{p}{)}\PY{p}{)}

\PY{k}{def} \PY{n+nf}{logsoftmax}\PY{p}{(}\PY{n}{logits}\PY{p}{)}\PY{p}{:}
  \PY{k}{return} \PY{n}{logits} \PY{o}{\PYZhy{}} \PY{n}{logsumexp}\PY{p}{(}\PY{n}{logits}\PY{p}{,} \PY{n}{axis}\PY{o}{=}\PY{o}{\PYZhy{}}\PY{l+m+mi}{1}\PY{p}{,} \PY{n}{keep\PYZus{}dims}\PY{o}{=}\PY{n+nb+bp}{True}\PY{p}{)}

\PY{k}{def} \PY{n+nf}{softmax\PYZus{}crossent}\PY{p}{(}\PY{n}{logits}\PY{p}{,} \PY{n}{y}\PY{p}{)}\PY{p}{:}
  \PY{k}{return} \PY{o}{\PYZhy{}}\PY{n}{np}\PY{o}{.}\PY{n}{sum}\PY{p}{(}\PY{n}{logsoftmax}\PY{p}{(}\PY{n}{logits}\PY{p}{)} \PY{o}{*} \PY{n}{y}\PY{p}{,} \PY{n}{axis}\PY{o}{=}\PY{o}{\PYZhy{}}\PY{l+m+mi}{1}\PY{p}{)}

\PY{k}{def} \PY{n+nf}{mlp}\PY{p}{(}\PY{n}{x}\PY{p}{,} \PY{n}{w1}\PY{p}{,} \PY{n}{b1}\PY{p}{,} \PY{n}{wout}\PY{p}{,} \PY{n}{bout}\PY{p}{,} \PY{n}{label}\PY{p}{)}\PY{p}{:}
  \PY{n}{h1} \PY{o}{=} \PY{n}{np}\PY{o}{.}\PY{n}{tanh}\PY{p}{(}\PY{n}{np}\PY{o}{.}\PY{n}{dot}\PY{p}{(}\PY{n}{x}\PY{p}{,} \PY{n}{w1}\PY{p}{)} \PY{o}{+} \PY{n}{b1}\PY{p}{)}
  \PY{n}{out} \PY{o}{=} \PY{n}{np}\PY{o}{.}\PY{n}{dot}\PY{p}{(}\PY{n}{h1}\PY{p}{,} \PY{n}{wout}\PY{p}{)} \PY{o}{+} \PY{n}{bout}
  \PY{n}{loss} \PY{o}{=} \PY{n}{np}\PY{o}{.}\PY{n}{mean}\PY{p}{(}\PY{n}{softmax\PYZus{}crossent}\PY{p}{(}\PY{n}{out}\PY{p}{,} \PY{n}{label}\PY{p}{)}\PY{p}{)}
  \PY{k}{return} \PY{n}{loss}

\PY{c+c1}{\PYZsh{} After generating these functions for calculating the derivative}
\PY{c+c1}{\PYZsh{} of \PYZsq{}mlp()\PYZsq{}, we timed their execution}
\PY{n}{autograd\PYZus{}dmlp} \PY{o}{=} \PY{n}{autograd}\PY{o}{.}\PY{n}{multigrad}\PY{p}{(}\PY{n}{mlp}\PY{p}{,}\PY{n}{argnums}\PY{o}{=}\PY{p}{(}\PY{l+m+mi}{1}\PY{p}{,}\PY{l+m+mi}{2}\PY{p}{,}\PY{l+m+mi}{3}\PY{p}{,}\PY{l+m+mi}{4}\PY{p}{)}\PY{p}{)}
\PY{n}{tangent\PYZus{}dmlp} \PY{o}{=} \PY{n}{tangent}\PY{o}{.}\PY{n}{grad}\PY{p}{(}\PY{n}{mlp}\PY{p}{,}\PY{n}{wrt}\PY{o}{=}\PY{p}{(}\PY{l+m+mi}{1}\PY{p}{,}\PY{l+m+mi}{2}\PY{p}{,}\PY{l+m+mi}{3}\PY{p}{,}\PY{l+m+mi}{4}\PY{p}{)}\PY{p}{)}
\end{Verbatim}

\caption{Multilayer perceptron used for benchmarking Tangent and Autograd}
\label{lst:mlp}
\end{listing}

\section{Conclusion}

We have introduced the AD library Tangent, highlighted several of its unique features, and compared
its performance to existing AD libraries.

Tangent is unique in that both the original function and the generated gradient
are pure Python, which allows for easy debugging and introspection. Gradient
transformation is a first class operation in Tangent which does not require
direct manipulation of the internal representation or the redefinition of
primitives. We believe that this enables easier development of complex machine
learning models in Python.

We have only described the use of Tangent with NumPy, but it is agnostic to the
numeric libraries used, as long as gradients for library functions are defined. We plan
to extend support of Tangent to other numeric libraries, particularly those with GPU support.
In future work, we hope to make it easier to express larger and more complicated
models in Tangent, as well as increase the subset of Python that Tangent supports.
We plan to release Tangent as a free and open-source library on GitHub in late November 2017.

\bibliographystyle{unsrt}
\bibliography{abstract}{}

\end{document}